# Stabilizing spin systems via symmetrically tailored RKKY interactions

Jan Hermenau[1,#], Sascha Brinker[2,#], Marco Marciani[3,ł], Manuel Steinbrecher[1], Manuel dos Santos Dias[2], Roland Wiesendanger[1], Samir Lounis[2] and Jens Wiebe[1,*]

[1] Department of Physics, Hamburg University, D-20355 Hamburg, Germany

[2] Peter Grünberg Institut and Institute for Advanced Simulation, Forschungszentrum Jülich &

JARA, 52425 Jülich, Germany

[3] Instituut-Lorentz, Universiteit Leiden, P.O. Box 9506, 2300 RA Leiden, The Netherlands

*Corresponding author: jwiebe@physnet.uni-hamburg.de

 łPresent address: Univ Lyon, Ens de Lyon, Univ Claude Bernard, CNRS, Laboratoire de Physique AF-69342 Lyon, France

#J. Hermenau and S. Brinker contributed equally to this work.

**The spin of a single atom adsorbed on a substrate is a promising building block for future spintronics and quantum computation schemes. To process spin information and also for increased magnetic stability, these building blocks have to be coupled. For a single atom, a high symmetry of the environment is known to lead to increased spin stability. However, little is known about the role of the nature and symmetry of the magnetic couplings. Here, we study arrays of atomic spins coupled via the ubiquitous Ruderman-Kittel-Kasuya-Yosida (RKKY) interaction, focusing on its two anisotropic parts: the Dzyaloshinskii-Moriya (DM) and the symmetric anisotropic exchange interactions. First, we show that the high spin stability of an iron trimer can be remotely detected by a nearby iron atom, and how the DM interaction can lead to its destabilization. Second, we find that adding more nearby iron atoms almost always leads to a destabilization of the trimer, due to a non-local effective transverse anisotropy originating in the symmetric anisotropic exchange interaction. This transverse anisotropy can be quenched only for highly symmetric structures, for which the spin lifetime of the array is increased by orders of magnitude.**

In magnetic systems, the enhancement, preservation or breaking of symmetries have strong implications for the ground-state and the dynamical properties. For instance, two-dimensional spin lattices with $C_{4V}$ symmetry and antiferromagnetic nearest-neighbor interactions form a stable diagonal row-wise antiferromagnet, while a $C_{6V}$ symmetry is predicted to have a spin liquid ground state [1-3]. The breaking of inversion symmetry is responsible for the presence of the Dzyaloshinskii-Moriya (DM) interaction between the spins in materials with large spin-orbit interaction [4, 5], which promotes the formation of magnetic skyrmions [6-9]. Another important example is the symmetry of the Coulomb field of the atoms around the spin, the so-called crystal field, which induces a magnetic anisotropy energy that leads to the stabilization of the spin along particular directions [10]. It has recently been shown that, when the crystal field of a single absorbed atomic spin has high symmetry [11], and the spin is decoupled from the conduction electrons of the substrate, the spin state lifetime can be enormously enhanced, up to hours [12]. This brings the use of such atomic spins as bits of information technology into reach.

In order to read, write and process the information stored in the state of such atomic spins, they have to be coupled to neighboring spins. The transfer of information can be realized via dipolar interaction [13], or the conduction-electron-mediated Ruderman-Kittel-Kasuya-Yosida (RKKY) interaction between spins on metallic surfaces [14-16]. The latter is unique in its flexibility to tailor the properties of the spin-system in terms of coupling strength and non-collinearity [17]. The general capability of RKKY-coupled networks to transfer and process information at the nanoscale has been shown by realizing few-atom spin-based logic gates [18]. One might naively expect that enhancing the number of atoms in such a network will drive the system into the regime of a classical magnet, where spin-fluctuations are less important. Indeed, exchange-coupling a few spins to larger ensembles via strong isotropic ferromagnetic (FM) [19] or antiferromagnetic (AFM) [20,21] Heisenberg exchange interactions has been shown to enhance the spin stability with respect to that of the constituents [22]. However, the coupling more generally has other contributions, such as the DM and the symmetric anisotropic Heisenberg exchange [23]. The effects of such contributions on the spin fluctuations and whether they are governed by symmetries remain to be explored. A first experimental and theoretical approach towards this end for RKKY-coupled spins is given in the following, where the spin-fluctuations of atom-by-atom engineered complexes consisting of one cluster and a different number of RKKY-coupled satellite atoms are investigated. The results of time-resolved spin-polarized scanning tunneling microscopy experiments are analyzed and interpreted taking into account the outcome of density functional theory (DFT) calculations and dynamic simulations.

## Results

**Trimer–satellite atom complexes**

A prototypical system for these studies consists of iron (Fe) atoms on a Pt(111) surface (Fig.1a), which adsorb either on the face-centered cubic (fcc) or on the hexagonal close-packed (hcp) sites. In both cases, the atoms can be approximately described by an effective spin with a quantum number of $S_a = 5/2$, and a preferred spin orientation along the surface normal (fcc, easy-axis anisotropy) or in the surface plane (hcp, easy-plane anisotropy) [24]. Up to distances of several lattice spacings, there is a sizeable RKKY interaction between the Fe atoms with significant Heisenberg as well as DM contributions. Varying the separation of the spins gives control over the magnitudes and signs of the Heisenberg and DM parts, in order to tailor noncollinear spin states [17, 25]. Contrary to the conventional wisdom, the strong interaction of Fe atoms with the Pt(111) substrate does not prevent ground state spin lifetimes of hours, when at least three iron atoms are manipulated into a close-packed fcc-top-stacked symmetrical trimer (Fig.1a) with a spin of $S_t = 11/2$ [26]. Since the spin state of these trimers (up or down, perpendicular to the surface) can be easily set by a short pulse of spin-polarized electrons with sufficient energy, they present ideal inputs/switches for RKKY-coupled networks of atoms [18]. However, it is a priori unclear whether the RKKY interaction, e.g. between such a trimer and a close-by satellite atom (Fig.1a), further stabilizes or destabilizes the spin state of the coupled - network. In order to investigate this question, we first place a single Fe atom on a fcc site at a distance of $d = 0.97$ nm to a well-characterized fcc top trimer [26], via lateral atom manipulation (Fig.1a).

**Two spin-state telegraph noise**

The strong impact of the RKKY coupling on the dynamics of the trimer-satellite atom system becomes apparent when a magnetic tip (sensitive to the out-of-plane component of the sample spin-polarization) is positioned on either of the two constituents of this complex (Fig.1b). For both tip positions, we observe a telegraph noise between two spin states, 0 and 1, with corresponding lifetimes of $\tau_0$ and $\tau_1$ and mean lifetime $\tau = (\tau_0^{-1} + \tau_1^{-1})^{-1}$ on the order of 10 ms to 10 s. Similar spin-state lifetimes have been observed for isolated uncoupled trimers and are associated with the tunneling electron induced dynamics between two degenerate out-of-plane oriented ground states of the trimer [26]. However, the observation of a lifetime on this order of magnitude for the spin state of the satellite atom is surprising. For isolated uncoupled atoms, the spin ground state lifetimes are at least one order of magnitude shorter [27], and the corresponding telegraph noise cannot be observed due to the limited time resolution of the method. Therefore, the telegraph noise detected on the satellite atom (Fig.1b bottom trace) must be a consequence of its RKKY interaction with the slowly switching trimer. We propose the mechanism illustrated in Fig.1c, which has been confirmed by our simulations (see below and Supplementary Notes 4 and 5). The RKKY interaction with the trimer induces an asymmetry $\frac{\tau_1^{a,short} - \tau_0^{a,short}}{\tau_1^{a,short} + \tau_0^{a,short}}$ in the short spin-ground-state lifetimes of the satellite atom ($\tau_{0/1}^{a,short}$). Although the SP-STM measurement averages out this fast telegraph noise, a changing asymmetry induced by switches of the trimer spin leads to a changing apparent height $z$ of the satellite atom, which is detectable via the longer lifetimes $\tau_0^a$ and $\tau_1^a$. Thereby, the telegraph noise which is measured on the satellite atom reflects only the slow dynamics of the coupled system and essentially enables a non-local read-out of the trimer spin similar to Ref. [20]. This interpretation is further corroborated by the correlation between the telegraph noises measured on the trimer and satellite atom (Fig.1d,e). In consecutively recorded line scans along the dashed line in Fig.1d, which are stacked on top of each other in Fig.1e, an anti-correlated switching of the trimer and satellite atom is found. Whenever the trimer is in the high state, the apparent height of the atom is low, and vice versa. Such anti-correlations or correlations of the two telegraph signals furthermore enable to conclude on a dominating AFM or FM orientation between the trimer and the satellite atom, since atom [28] and trimer possess the same sign of the vacuum spin-polarization (Supplementary Note 2). The anti-correlation observed in Fig.1e therefore indicates an AFM spin orientation in this particular complex.

**Dependence of the dynamics on the separation of trimer and satellite atom**

Although the telegraph noises measured in the two situations depicted in the insets in Fig.1b, i.e. directly on the trimer (black) or non-locally on the satellite atom (red), reflect the dynamics of the coupled system, they strongly differ for the two cases. Consequently, these differences must originate from different probabilities to switch the trimer-satellite atom complex for the two tunneling paths. Interestingly, for the particular complex in Fig.1, the tunnel current is more likely to induce switching via the atom than via the trimer ($\tau_a < \tau_t$), and both lifetimes are smaller than the lifetime measured on the same trimer in the absence of the satellite atom [26]. Counter-intuitively, the trimer spin is found to be destabilized by the nearby atom.

In order to systematically study the influence of the satellite atom on the magnetization dynamics of the complex, various structures are constructed and characterized in terms of dynamics and coupling behavior. To this end, the satellite atom is moved to adsorption sites at various distances to the trimer via lateral atom manipulation (Fig.2a). In all these complexes, a finite coupling was revealed by a telegraph noise on the satellite atom analogous to the observations for position "a" in Fig.1. In dependence of the actual position of the satellite atom, the telegraph noises of trimer and atom are correlated or anti-correlated, indicating a complex oscillation between FM and AFM orientation illustrated in Fig.2a by the corresponding color. This oscillatory behavior is reminiscent of an RKKY interaction. Indeed, the measured orientations are consistent with the configurations that have been obtained from the minimization of a classical Heisenberg model using density DFT parameters as input (Fig.2b, see Methods). The DFT calculations were performed for all fcc positions of the satellite atoms under the assumption of a collinear internal spin configuration in the out-of-plane direction for the trimer, which is a valid approximation for fcc trimers due to only tiny noncollinear deviations [26]. The calculations reveal that the magnetic moment of the satellite atom is almost collinear to that of the trimer (along the $z$-axis, which is the easy axis for fcc atoms). Only for the farthest separation between trimer and satellite atom, are the experimental and theoretical results in contradiction, which might be due to inaccuracies in the calculation associated with the small size of the interaction. The FM-AFM oscillation results from a complex interplay of different contributions to the RKKY interaction, as revealed by mapping the DFT calculated energies to a generalized classical Heisenberg model which then results in the quantum Heisenberg Hamiltonian

$$\hat{H} = J_{\mathrm{iso}}\hat{\mathbf{S}}_\mathrm{a} \cdot \hat{\mathbf{S}}_\mathrm{t} + \hat{\mathbf{S}}_\mathrm{a} \cdot \underline{\mathbf{J}}_{\mathrm{aniso}} \cdot \hat{\mathbf{S}}_\mathrm{t} + \mathbf{D} \cdot \hat{\mathbf{S}}_\mathrm{a} \times \hat{\mathbf{S}}_\mathrm{t} + \mathcal{D}_a\big(\hat{S}_\mathrm{a}^z\big)^2 + \mathcal{D}_t\big(\hat{S}_\mathrm{t}^z\big)^2, \qquad (1)$$

with the vector spin operators of the satellite atom $\hat{\mathbf{S}}_\mathrm{a}$ and trimer $\hat{\mathbf{S}}_\mathrm{t}$ and their respective $z$-components $\hat{S}_\mathrm{a}^z$ and $\hat{S}_\mathrm{t}^z$. Besides the magneto-crystalline anisotropies (MAE) of atom ($\mathcal{D}_a$) and trimer ($\mathcal{D}_t$), it includes the most general form of Heisenberg exchange including the isotropic ($J_{\mathrm{iso}}$) and the symmetric anisotropic ($\underline{\mathbf{J}}_{\mathrm{aniso}}$) parts, as well as the DM interaction (**D**) (see Supplementary Note 3,4 for the definition of the operators and the relation between classical and quantum Heisenberg Hamiltonian). The symmetric anisotropic part, $\underline{\mathbf{J}}_{\mathrm{aniso}}$, is also called the pseudo-dipolar interaction or a two-ion anisotropy term, which contrary to the isotropic part, $J_{\mathrm{iso}}$, favors the orientation of the moments along specific directions similarly to MAE, leading to non-local effective magnetic anisotropies.

Fig.2c shows the distance-dependent strengths of $J_{\mathrm{iso}}$, $\underline{\mathbf{J}}_{\mathrm{aniso}}$ and **D**, whose components all display oscillations in the sign typical for RKKY interaction (see Fig.2d). These sign oscillations, together with a rapid decrease in the strength of the RKKY interaction (Fig.2c) and the relatively strong out-of-plane MAE of the fcc satellite atoms $\mathcal{D}_\mathrm{a} = -0.19\ \mathrm{meV}$ [24] and trimer $\mathcal{D}_\mathrm{t} =$ $-0.09\,\mathrm{meV}$ [26] lead to an almost collinear spin orientation between trimer and satellite atom along the surface normal, which alternates between FM and AFM as their separation is varied. For complexes including hcp satellites the experimental results (Supplementary Note 1) point towards a deviation of the atom's spin orientation from the surface normal due to the in-plane MAE.

In order to investigate the influence of these RKKY couplings on the dynamics of the complexes, time traces like those in Fig.1b were recorded for each built configuration. The lifetimes $\tau_\mathrm{a}$ and $\tau_\mathrm{t}$ measured with the tip on the satellite atom, respectively the trimer, are plotted in Fig.3a against the atom-trimer separation. Obviously, there are two qualitatively different regimes. For smaller distances ($d < 1.25\,\mathrm{nm}$, see vertical dashed line), there is a destabilization of the trimer by the satellite atom in comparison to the free-standing case (horizontal dashed line). Independently of whether the complex lifetime is measured via the atom (red) or directly on the trimer (black), the values are about one order of magnitude smaller. In contrast, for larger distances ($d > 1.25\,\mathrm{nm}$), the lifetime of the complex strongly differs between the two measurement modes: When the current runs through the trimer ($\tau_\mathrm{t}$) the lifetime is comparable to that of an isolated trimer, which indicates a negligible influence of the satellite atom on the trimer dynamics. However, the two-state telegraph noise can still be observed if the current runs through the satellite atom ($\tau_\mathrm{a}$), proving a finite RKKY coupling between the two constituents. The accordingly-measured values of $\tau_\mathrm{a}$ are two orders of magnitude larger than $\tau_\mathrm{t}$, indicating that the influence of the measurement process on the trimer dynamics is drastically reduced. No clear correlation is found between the lifetime and the type of the adsorption site of the satellite atom nor the type of coupling, AFM or FM. Also, the destabilization effect in the small distance regime is independent of the used bias voltage, as it becomes obvious by comparison of the data from the free-standing trimer and the complex recorded over a large range of bias voltages (Fig.4l).

To unravel the physics behind the sharp crossover between the two regimes, we performed simulations within a master equation approach, in conjunction with the effective quantum spin Hamiltonian given above [19, 26] using the exchange parameters extracted from the DFT calculations (see Methods). In both regimes, the simulations reproduce the experimental data quantitatively (Fig.3a). Only for larger distances and when the current flows through the satellite atom, does the model overestimate the lifetimes, which might be explained by quantum fluctuations due to additional transversal magnetic anisotropies [27] or zero-point spin fluctuations [29] which go beyond those included in the master equation approach.

The simulations enable to identify the origin of the steep increase of the lifetimes around $1.25$ nm by independently sweeping $D_x$ and $J_{\text{iso}}$ (Fig.3c). In the experimentally relevant parameter regime of atom-trimer separations of around $1.25$ nm, there is a strong drop of the lifetime when the DM interaction strength exceeds $\approx 6 \cdot 10^{-3}$ meV. The effect of the isotropic exchange interaction on the lifetime is much less pronounced. This strongly points towards a dominant destabilization of the trimer spin by the DM interaction. Further studies separating the dependence of the lifetime on the three different exchange contributions reinforce this finding (Fig.3b). Contrary to the DM interaction, the symmetric anisotropic exchange interaction can affect the lifetimes in two ways: It can either stabilize the complex by inducing an effective easy-axis anisotropy (e.g. via a ferromagnetic *zz*-component) or destabilize it, similarly to the DM interaction, by inducing an effective transversal anisotropy. Fig.3b shows that the destabilization of the lifetime for an atom-trimer separation of around $1.25$ nm is mainly driven by the DM interaction, whereas in the close distance regime the symmetric anisotropic part of the exchange interaction dominates the destabilization.

**Dependence of the dynamics on symmetry**

The observed destabilization of the spin-structure by coupling the trimer to a satellite atom is rather counterintuitive, as generally systems tend to stabilize with an increasing number of spins [19, 21]. Importantly, the addition of the satellite atom not only increases the number of spins, but also lowers the symmetry of the system. These two effects on the lifetime, i.e. number of spins and symmetry, are investigated in the following by adding different numbers of satellite atoms in different configurations to the complex (Fig.4a-j). The lifetime depends in a rather intricate manner on the number of satellite atoms. It first increases for two satellite atoms (Fig.4c,l), but then stays almost the same for three (e) and four (f) satellite atoms, or is even drastically reduced for a different position of the four satellite atoms (g). Note, that from (b) to (g) all satellite atoms have a distance $d < 1.25$ nm to the trimer, i.e. within the regime of strong coupling (cf. Fig3a). In most of these complexes the lifetime is smaller than for the symmetrical case of the trimer without any satellite atoms (a). Only if the satellite atoms are added in a way that conserves the $C_{3V}$ symmetry of the trimer/substrate (d), can the lifetime be further increased by an order of magnitude above the one of the isolated trimer.

This suggests that keeping a high symmetry of the RKKY interactions, and in particular of the symmetric anisotropic contribution which dominates the destabilization in the complex with one satellite atom in the close distance regime, restores the spin-stability. The effect of the symmetry of the RKKY interactions on the lifetime can be rationalized by the DFT calculation of the effective magnetic anisotropy energy of all trimer-satellite atom complexes when the spins of all constituents are collinearly rotated in the surface plane (Fig.4m). Interestingly, the RKKY interaction induces an effective non-local transversal magnetic anisotropy in the complex spin system, whenever the $C_{3V}$ symmetry is broken resulting in $C_S$ or no symmetry. On-site transversal magnetic anisotropy in single-spin effective Hamiltonians is well known to induce instability of the lifetime via mixing of the spin states [30, 31]. However, the related effect in a multi-spin complex originating from the non-local RKKY interactions between the spins, which we observe here, has not been studied so far, and cannot be reduced to the single-spin effective Hamiltonian picture.

We consequently expect that it might be possible to further stabilize the complex by adding a larger number of satellite atoms in symmetrical positions. This idea is experimentally verified by the formation of a symmetrical complex with six satellites shown in Fig.4h. Successively adding three more atoms on symmetrical positions to the complex in (d), the lifetime is further increased by more than an order of magnitude (see the intermediate steps in (i,j) and the corresponding lifetimes in Fig.4l).

**Conclusions**

In conclusion, we revealed the dynamic behavior of RKKY-coupled complexes of two and more constituents. There are three regimes depending on the coupling strength between the trimer and a single satellite atom: In the weak coupling regime, the nearly-intrinsic dynamics of the trimer could be observed while probing the satellite atom. This enables to non-locally read and write the trimer spin via an RKKY coupled network of atoms. In the intermediate coupling regime, the DM interaction leads to a significant destabilization when its magnitude exceeds a certain threshold. In the strong coupling regime, the symmetric anisotropic exchange interaction leads to a non-local transverse effective magnetic anisotropy which generally destabilizes the system. However, in the latter case, by engineering the symmetry of the complex, the magnetic stability can be restored and even enhanced by an order of magnitude. This demonstrates a new direction for the tailoring of RKKY coupled networks of atoms towards usage in spintronics elements.

## Methods

**Experimental procedures**

All measurements have been performed under ultra-high vacuum conditions in a home-built scanning tunneling microscope facility at a base temperature of $T = 300\,\text{mK}$. A superconducting magnet enabled to apply a magnetic field $B$ perpendicular to the sample surface [32]. The Pt(111) single crystal was cleaned in situ by argon ion sputtering and annealing cycles and a final flash as described in Ref. [24]. The experiments described here have been conducted using the same preparation as used for the measurements presented in Refs. [25-27].

The trimer was constructed by lateral atom manipulation [33] in the pulling mode. The construction and identification is described in detail in Ref. [26]. With the same technique, the arrangement of satellite atoms in the complexes is changed. During such a rearrangement of the complex, residual vertical or lateral drift is compensated. This enables to exclude tip changes by comparing the absolut substrate height before and after each manipulation. The stacking of the new position of the satellite atom is revealed by inelastic scanning tunneling spectroscopy [24]. The trimer is not moved within the presented experiments.

All measurements presented here are performed using a spin-polarized tip with a magnetically stable sensitivity to the out-of-plane component of the sample magnetization. To this end, the spin sensitivity of a Cr coated tungsten tip was optimized by picking up single iron atoms until a high two-state telegraph noise is observed on the trimer.

Constant-current images were taken using a tunnelling current $I$ with a bias voltage $V$ applied to the sample. For the dynamic measurements, the tip is placed stationary above the atom or the trimer and the apparent height is recorded with running feedback loop stabilizing the current at $I$.

**Ab initio calculations**

For the ab initio density functional theory (DFT) calculations, we employed the Korringa-Kohn-Rostoker Green function method in full potential with spin-orbit coupling added to the scalar-relativistic approximation [34]. The exchange and correlation effects are treated in the local spin density approximation (LSDA) as parametrized by Vosko, Wilk and Nusair [35]. The pristine Pt(111) surface is modeled by 40 Pt layers augmented by two vacuum regions. The magnetic Fe atoms are placed on the surface in the fcc-stacking position, using an embedding method. The magnetic anisotropy energies and the exchange interactions are obtained from band energy differences using the magnetic force theorem [36] and the infinitesimal rotation method [36, 37], respectively. Both are used in an extended classical Heisenberg model to obtain the ground state spin configuration (see Supplementary Note 3). The DFT results for the spin moments of the iron atoms and their magnetic anisotropy and magnetic pair interactions were used to parametrize the master equation model.

**Master equation model**

We use a master equation model to simulate the lifetime in the telegraph noise experiment. The magnetic clusters are described within the quantum Heisenberg model (equ. 1) following Ref. [19, 26]. The interaction of the tunneling electrons with the spins is described by an Applebaum Hamiltonian [38]. The ratio between the coupling of the spins to the tip and to the surface at a bias voltage of 5 V is set to $v^{\text{T}}/v^{\text{S}} = 0.055$, as determined in a previous experiment [26]. The lifetimes are determined using a master equation approach combined with Fermi's Golden rule [38].

**Data availability**

The authors declare that the main data supporting the findings of this study are available within the article and its Supplementary Information files. Extra data are available from the corresponding author upon reasonable request.

# Acknowledgements

J.H., M.S., R.W., and J.W. acknowledge funding from the SFB668 and the GrK1286 of the DFG. S.B., M.d.S.D. and S.L. acknowledge funding from the European Research Council (ERC) under the European Union's Horizon 2020 research and innovation program (ERC-consolidator grant 681405 – DYNASORE) and the computing time granted by the JARA-HPC Vergabegremium and VSR commission on the supercomputer JURECA at Forschungszentrum Jülich. M.M. was supported by the Netherlands Organization for Scientific Research (NWO/OCW), an ERC Synergy Grant and the French Agence Nationale de la Recherche (ANR) under Grant TopoDyn (ANR-14-ACHN-0031).

# Author contributions

J.H. and J.W. designed the experiments. J.H. carried out the measurements together with M.S. and did the analysis of the experimental data. S.B. performed the DFT calculations. M.M., S.B. and M.d.S.D. extended the master equation model, which S.B. used to simulate the data. J.H. and J.W. wrote the manuscript, to which all authors contributed via discussions and corrections.

# Competing financial interests

The authors declare no competing financial interests.

# Figures

**Figure 1 | Assembly and spin switching of the trimer – satellite atom complex. a** Constant-current STM images ($5x5$ nm$^2$, $V = 5$ mV, $I = 500$ pA) of the construction process of a complex via lateral atom manipulation ($V_{\mathrm{manip}} = 1.1$ mV, $I_{\mathrm{manip}} = 25$ nA). A single iron atom is approached to the trimer. Within the complex the atom sits on a fcc adsorption side labelled as "a" in Fig. 2a. In this complex, the distance between atom and trimer is $d = 0.97$ nm. **b** Time traces of the apparent height recorded on the trimer (black) and on the atom (red) ($V = 2.75$ mV, $I = 200$ pA, $B = 0$ T, $T = 300$ mK). The data sets are shifted vertically for better visibility. **c** Illustration of the contrast mechanism for the data taken on the fast switching atom ferromagnetically coupled to a trimer in a SP-STM measurement. The fast switching events are averaged out during the measuring time, leading to a change in the apparent height of the atom when the trimer switches as indicated by the dark red lines. **d** Constant-current STM image of the complex recorded with an out-of-plane spin-polarized tip ($V = 3$ mV, $I = 50$ pA, $B = 0$ T, $T = 300$ mK). **e** Consecutively recorded spin-resolved line scans along the dashed line in **d** stacked on top of each other ($V = 3$ mV, $I = 20$ pA, $B = 0$ T, $T = 300$ mK, scanning time per line = 614 ms).

**Figure 2 | Distance dependence of the spin configuration and the coupling strength. a**, experimentally investigated atom positions labelled by the chronological order they were addressed in the experiment. Fcc (hcp) positions are marked with circles (stars). The color depicts whether a FM (purple) or an AFM (green) switching was revealed experimentally as the one shown for position "a" in Fig.1 d,e. For the positions shown in grey, no contrast/switching on the atom was found. Three exemplary complex configurations are shown in the constant-current images on the left. **b** Ground state spin configurations for fcc adsorbed satellite atoms determined from a classical Heisenberg model using parameters obtained from DFT calculations. The trimer atoms sit on the three fcc sites in the center (white). $\vartheta$ describes the angle between the atom and the trimer spin. The fcc positions addressed in the experiment are marked by circles (the position which shows a mismatch between experiment and DFT is given in red). **c** Strengths $|J_{\mathrm{iso}}|$ (purple), $|\mathbf{J}_{\mathrm{aniso}}|$ (orange) and $|\mathbf{D}|$ (blue) in dependence of the distance between the trimer and the atom (see Supplementary Note 6 for the definition of the strengths). **d** $J_{\mathrm{iso}}$ (purple), $J_{\mathrm{xx}}$ (orange) and $D_x$ (blue) in dependence of the trimer-atom separation.

**Figure 3 | Distance dependence of the spin lifetime. a** Mean lifetimes of the data recorded on the trimer $\tau_t$ (black) and on the atom $\tau_a$ (red) in dependence of the trimer-atom separation ($V = -5$ mV, $I = 500$ pA, $B = 0$ T and $T = 300$ mK). Data points of the complexes with an fcc (hcp) adsorbed satellite atom are marked with a circle (star). The horizontal (vertical) dashed line marks the lifetime of the free-standing trimer (the crossover between the two lifetime regimes). In grey (weak red) the simulated lifetimes for the scenario of the tip probing the trimer (the satellite atom) are given. **b** Calculated lifetimes probed on the trimer for the experimental separations using different parts of the exchange interactions. The reference data using the full exchange interactions is shown in grey. The lifetimes calculated without the symmetric anisotropic exchange are shown in green. The lifetimes calculated without the DM interaction are shown in purple. **c** Calculated lifetimes probed on the trimer in dependence of the isotropic exchange interaction $J_{iso}$ (AFM: green, FM: purple) and the DM interaction $D_x$ (blue). For the calculations,

the two remaining exchange contributions are set to zero (Supplementary Note 5). The grey area marks the regime of the exchange constants found for an atom-trimer separation of about 1.25 nm.

**Figure 4 | Magnetic stability of multi-satellite complexes. a-j** Constant current images of the investigated complexes. For **b**-**d** the distance between satellite atom and closest trimer atom is $d = 0.8$ nm. **k** The corresponding adsorption sites of the complexes shown in **a-j**. **l** Bias dependence of the lifetimes measured on the trimer of the structures marked in **a-j** ($I = 750$ pA $B = 0$ T and $T = 300$ mK, colors/symbols are given in **a-j**). **m** Effective magnetic anisotropy energy of the spin system in dependence of a collinear rotation of the trimer and adatom spins in the sample surface (colors/symbols are given in **a-j**, 0° parallel to x-axis).

References

1. Anderson, P. W. Resonating valence bonds: A new kind of insulator? *Materials Research Bulletin* **8**, 153-160 (1973).
2. Normand, B. Frontiers in frustrated magnetism. *Contemporary Physics* **50**, 533-552 (2009).
3. Balents, L. Spin liquids in frustrated magnets. *Nature* **464**, 199 (2010).
4. Dzyaloshinsky, I. A thermodynamic theory of "weak" ferromagnetism of antiferromagnetics. *Journal of Physics and Chemistry of Solids* **4**, 241-255 (1958).
5. Moriya, T. Anisotropic superexchange interaction and weak ferromagnetism. *Physical Review* **120**, 91-98 (1960).
6. von Bergmann, K. *et al.* Observation of a complex nanoscale magnetic structure in a hexagonal Fe monolayer. *Physical Review Letters* **96**, 167203 (2006).
7. Bode, M. *et al.* Chiral magnetic order at surfaces driven by inversion asymmetry. *Nature* **447**, 190 (2007).
8. Heide, M., Bihlmayer, G. & Blügel, S. Dzyaloshinskii-moriya interaction accounting for the orientation of magnetic domains in ultrathin films: Fe/w(110). *Physical Review B* **78**, 140403 (2008).
9. Heinze, S. *et al.* Spontaneous atomic-scale magnetic skyrmion lattice in two dimensions. *Nature Physics* **7**, 713 (2011).
10. Bloch, F. & Gentile, G. Zur Anisotropie der Magnetisierung ferromagnetischer Einkristalle. *Zeitschrift für Physik* **70**, 395-408 (1931).
11. Rau, I. G. *et al.* Reaching the magnetic anisotropy limit of a 3d metal atom. *Science* **344**, 988-992 (2014).
12. Donati, F. *et al.* Magnetic remanence in single atoms. *Science* **352**, 318-321 (2016).
13. Natterer, F. D. *et al.* Reading and writing single-atom magnets. *Nature* **543**, 226 (2017).
14. Ruderman, M. A. & Kittel, C. Indirect exchange coupling of nuclear magnetic moments by conduction electrons. *Physical Review* **96**, 99-102 (1954).
15. Kasuya, T. A theory of metallic ferro- and antiferromagnetism on zener's model. *Progress of Theoretical Physics* **16**, 45-57 (1956).
16. Yosida, K. Magnetic properties of cu-mn alloys. *Physical Review* **106**, 893-898 (1957).
17. Steinbrecher, M. *et al.* Non-collinear spin states in bottom-up fabricated atomic chains. *Nature Communications* **9**, 2853 (2018).
18. Khajetoorians, A. A., Wiebe, J., Chilian, B. & Wiesendanger, R. Realizing all-spin–based logic operations atom by atom. *Science* **332**, 1062-1064 (2011).
19. Khajetoorians, A. A. *et al.* Current-driven spin dynamics of artificially constructed quantum magnets. *Science* **339**, 55-59 (2013).

20. Yan, S. *et al.* Nonlocally sensing the magnetic states of nanoscale antiferromagnets with an atomic spin sensor. *Science Advances* **3**(2017).
21. Loth, S., Baumann, S., Lutz, C. P., Eigler, D. M. & Heinrich, A. J. Bistability in atomic-scale antiferromagnets. *Science* **335**, 196-199 (2012).
22. Marciani, M., Hübner, C. & Baxevanis, B. General scheme for stable single and multiatom nanomagnets according to symmetry selection rules. *Physical Review B* **95**, 125433 (2017).
23. Juba, B. *et al.* Chiral magnetism of magnetic adatoms generated by rashba electrons. *New Journal of Physics* **19**, 023010 (2017).
24. Khajetoorians, A. A. *et al.* Spin excitations of individual Fe atoms on Pt(111): Impact of the site-dependent giant substrate polarization. *Physical Review Letters* **111**, 157204 (2013).
25. Khajetoorians, A. A. *et al.* Tailoring the chiral magnetic interaction between two individual atoms. *Nature Communications* **7**, 10620 (2016).
26. Hermenau, J. *et al.* A gateway towards non-collinear spin processing using three-atom magnets with strong substrate coupling. *Nature Communications* **8**, 642 (2017).
27. Hermenau, J., Ternes, M., Steinbrecher, M., Wiesendanger, R. & Wiebe, J. Long spin-relaxation times in a transition-metal atom in direct contact to a metal substrate. *Nano Letters* **18**, 1978-1983 (2018).
28. Steinbrecher, M. *et al.* Absence of a spin-signature from a single Ho adatom as probed by spin-sensitive tunneling. *Nature Communications* **7**, 10454 (2016).
29. Ibañez-Azpiroz, J., dos Santos Dias, M., Blügel, S. & Lounis, S. Zero-point spin-fluctuations of single adatoms. *Nano Letters* **16**, 4305-4311 (2016).
30. Delgado, F. & Fernández-Rossier, J. Spin decoherence of magnetic atoms on surfaces. *Progress in Surface Science* **92**, 40-82 (2017).
31. Hübner, C., Baxevanis, B., Khajetoorians, A. A. & Pfannkuche, D. Symmetry effects on the spin switching of adatoms. *Physical Review B* **90**, 155134 (2014).
32. Wiebe, J. *et al.* A 300 mk ultra-high vacuum scanning tunneling microscope for spin-resolved spectroscopy at high energy resolution. *Review of Scientific Instruments* **75**, 4871-4879 (2004).
33. Eigler, D. M. & Schweizer, E. K. Positioning single atoms with a scanning tunnelling microscope. *Nature* **344**, 524-526 (1990).
34. Bauer, D. S. G., Development of a relativistic full-potential first-principles multiple scattering green function method applied to complex magnetic textures of nano structures at surfaces, RWTH Aachen University
35. Vosko S. H, L., Wilk, L. H. & Nusair, M. *Accurate spin-dependent electron liquid correlation energies for local spin density calculations: A critical analysis* (1980)
36. Liechtenstein, A. I., Katsnelson, M. I., Antropov, V. P. & Gubanov, V. A. Local spin density functional approach to the theory of exchange interactions in ferromagnetic metals and alloys. *Journal of Magnetism and Magnetic Materials* **67**, 65-74 (1987).
37. Mankovsky, S. *et al.* Effects of spin-orbit coupling on the spin structure of deposited transition-metal clusters. *Physical Review B* **80**, 014422 (2009).
38. Delgado, F. & Fernández-Rossier, J. Spin dynamics of current-driven single magnetic adatoms and molecules. *Physical Review B* **82**, 134414 (2010).

Figure 1

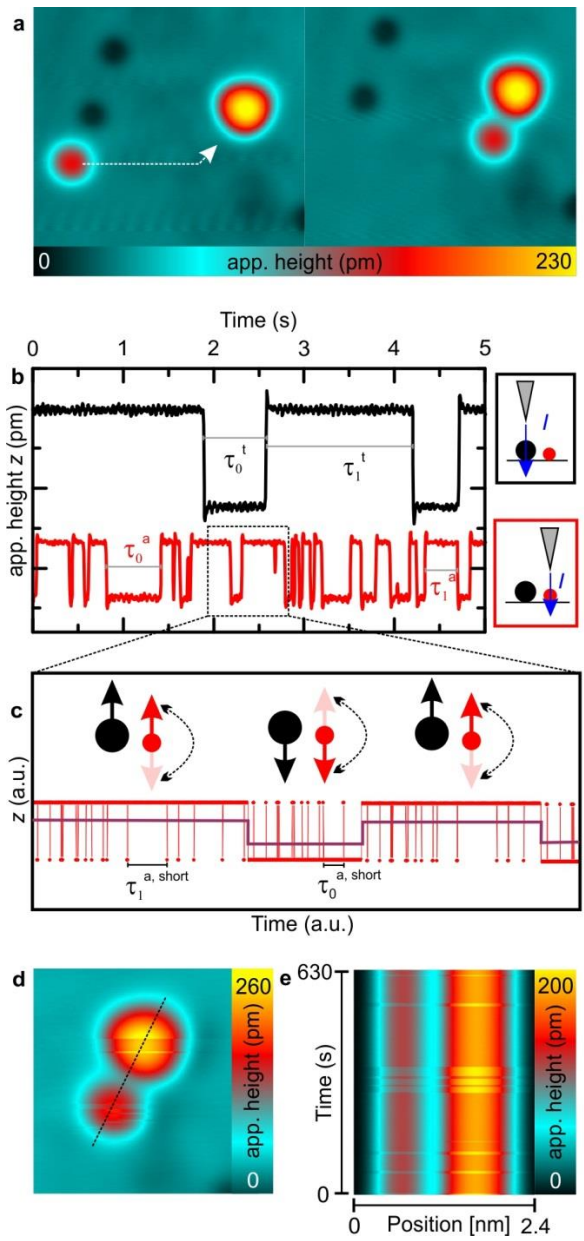

Figure 2

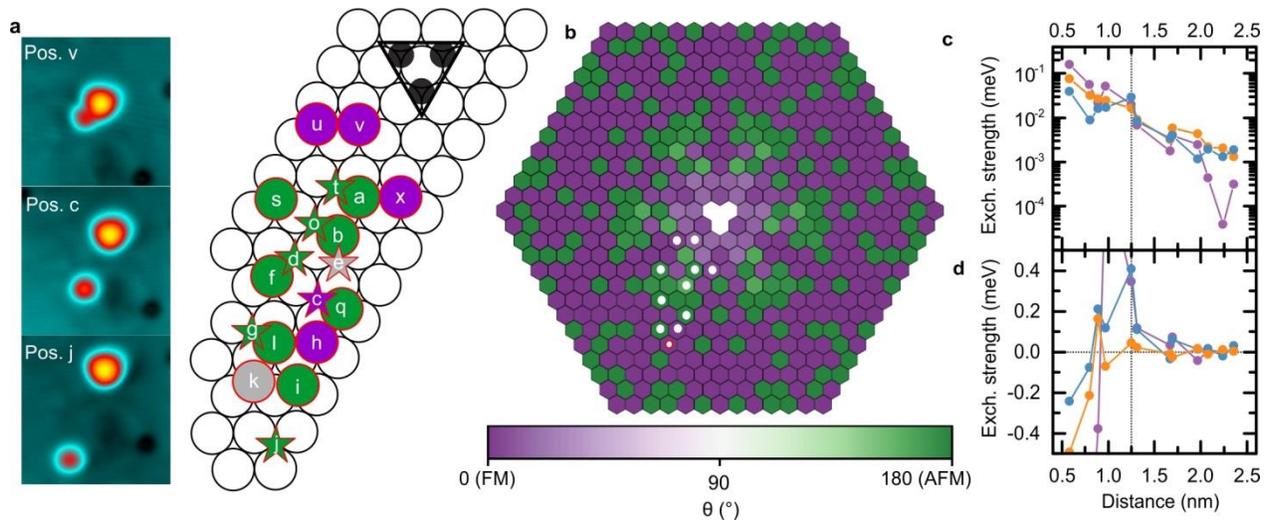

Figure 3

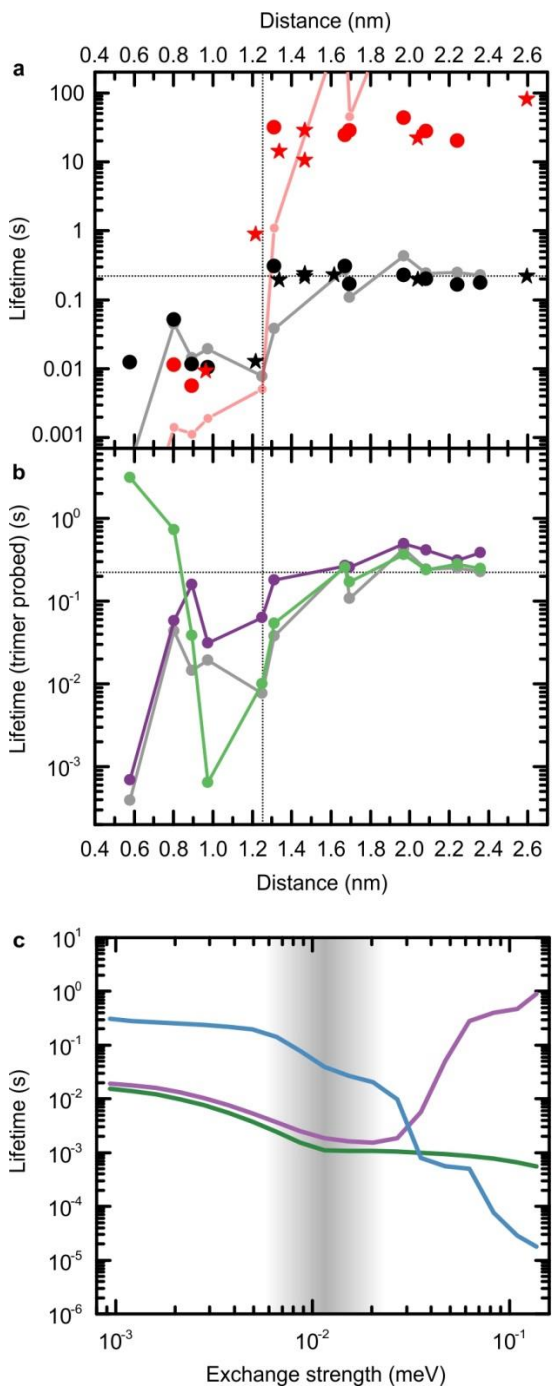

Figure 4

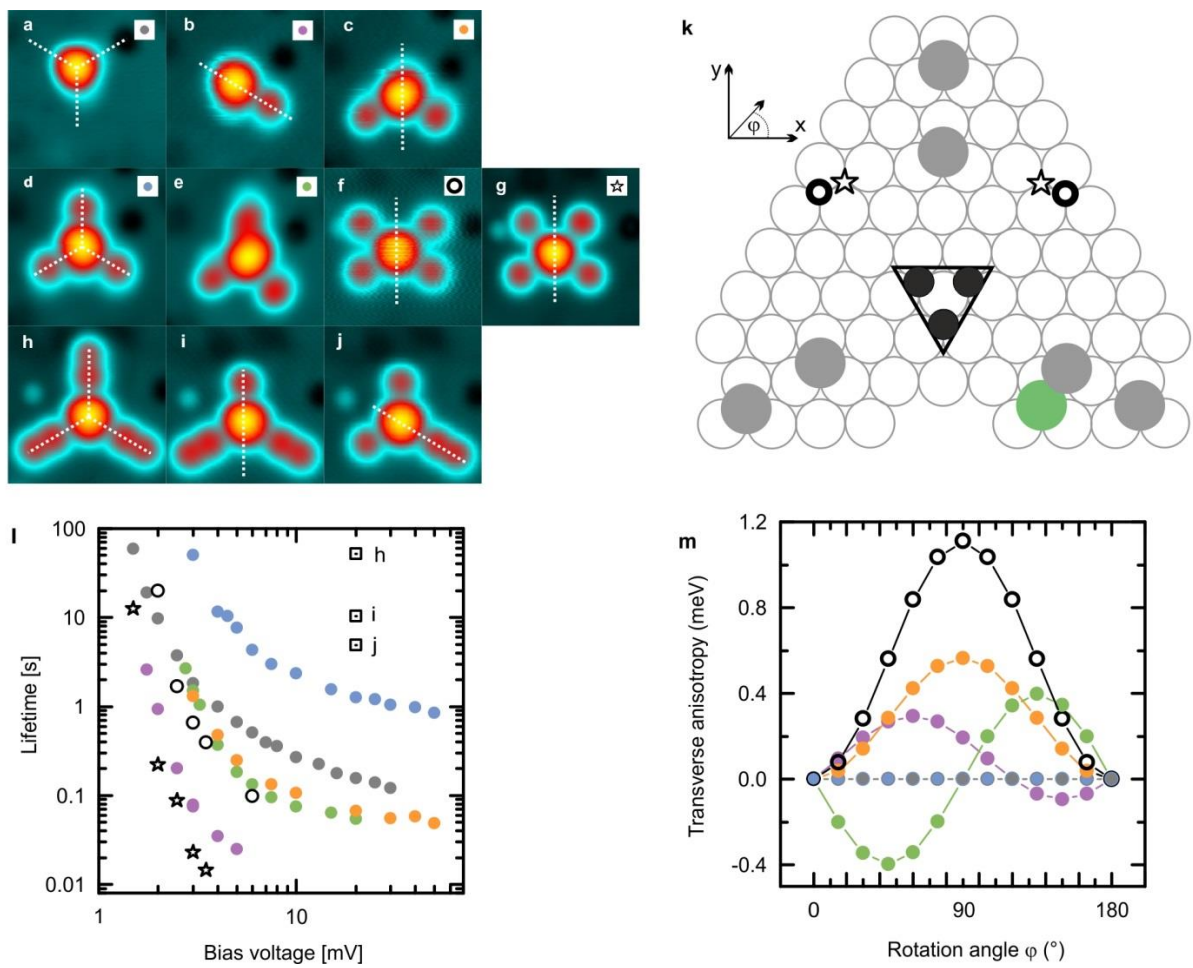